\documentclass[twocolumn,prl]{revtex4}

\begin{document}

\title{On the Electric Charge Quantization from the Dirac-Aharonov-Bohm Potential}

\author{F. A. Barone\footnote{e-mail:fbarone@cbpf.br} and J. A. Helay\"el-Neto\footnote{e-mail:helayel@cbpf.br}}
\affiliation{Centro Brasileiro de Pesquisas F\'\i sicas, Rua Dr.\ Xavier Sigaud 150, Urca 22290-180, Rio de Janeiro, RJ, Brazil}

\date{\today}

\begin{abstract}
	The purpose of this paper is to show that, under certain restrictions, we can take a  Dirac-Aharonov-Bohm potential as a pure gauge field. We argue that a modified quantization condition comes out for the electric charge that may open up the way for the understanding of fractional charges. One does not need any longer to rely on the existence of a magnetic monopole to justify electric charge quantization.
\end{abstract}

\maketitle

	The main goal of this paper is to show that, even in a quantum-mechanical context, we can take a Dirac-Aharonov-Bohm-like potential as a gauge field, once we consider some restrictions on these gauge potentials, and, as a consequance, we are led to a quantization condition for the electric charge which does not involve the existence of magnetic monopoles and which is in agreement with the fractional character of quark charges and the existence of anti-particles.

	Before start discussing electric charge quantization with no appeal to monopoles, let us briefly recall the main aspects of Dirac's quantization, which intrinsically involves the existence of magnetic monopoles \cite{Dirac}. Let us discuss this last point following the Wu and Yang approach \cite{WuYang}.

	To describe the field strength of a point-like magnetic monopole $g$, Dirac used the vector potential of a so-called Dirac's string \cite{Dirac,WuYang,GoddardOliverRPP78}, which, if considered to be lying along the $z<0$ semi-axis, gives the vector potential
\begin{equation}
\label{corda1}
{\bf A}^{I}=g\frac{1-\cos\theta}{r\sin\theta}{\bf\hat\phi}\ ,
\end{equation}
where we used spherical coordinates where $\phi$ is the azimuthal angle. With this description, Dirac avoided the introduction of a scalar potential of magnetic nature and, consequently, the trouble of establishing an electrodynamics with a pair of four potentials \cite{MignacoBJP01}. Even so, expression (\ref{corda1}) exhibits the problem of not reproducing the field strength of the magnetic monopole in the whole space; in fact \cite{GoddardOliverRPP78,Felsager}
\begin{equation}
{\bf\nabla}\times{\bf A}^{I}=\frac{g}{r^{2}}{\bf\hat r}+4\pi g\Theta(-z)\delta(x)\delta(y){\bf\hat z}\ ,
\end{equation}
which encompasses the field of the magnetic monopole plus a singular one along the string (in the above expression $\delta(x)$ is the Dirac delta function and $\Theta(\xi)$ is the step function which, as usually adopted, is non-defined for $\xi=0$).

	In the classical context, this is not actually a problem. As stated by Wu and Yand \cite{WuYang}, the potential (\ref{corda1}) must be defined only on a patch, which is the open set composed by the whole space except the $z<0$ semi-axis ($0\leq\theta<\pi$).
	
	By the same means, in order to describe the magnetic field in the $z<0$ semi-axis, we can take the string, for instance, on the $z>0$ semi-axis, which gives the potential
\begin{equation}
\label{corda2}
{\bf A}^{II}=-g\frac{1+\cos\theta}{r\sin\theta}{\bf\hat\phi}\ ,
\end{equation}
that must not be considered along $z>0$ semi-axis, it is, the potential (\ref{corda2}) must be defined only on the patch which excludes the $z>0$ semi-axis.

	There is an overlap region between the patches cited above where booth potentials (\ref{corda1}) and (\ref{corda2}) are defined. This overlap region is the open set composed by the whole space excluding the $z$ axis. As stated by Wu and Yang \cite{WuYang}, in this region the potentials (\ref{corda1}) and (\ref{corda2}) must be related by the singular gauge transformation \cite{WuYang}
\begin{equation}
\label{calibreDirac}
{\bf A}^{I}-{\bf A}^{II}=2g\frac{1}{r\sin\theta}{\bf\hat\phi}={\bf\nabla}\bigl({2g\phi}\bigr)
\end{equation}
which is not defined along the $z$ axis. With this approach, the string can be seen as a purely mathematical artifact, valid only in the classical context.

	It can be shown that the freedom of describing the magnetic monopole field using string fields, placed in arbitrary regions, still remains in the quantum context only if the so-called Dirac's quantization \cite{Dirac,WuYang,GoddardOliverRPP78,MignacoBJP01}
\begin{equation}
\label{quantizacaoDirac}
qg=\frac{\hbar c}{2}n
\end{equation}
is valid, where $n$ is an integer and $q$ stands for the possible electric charges a wave function may have. If the condition (\ref{quantizacaoDirac}) was not satisfied, the strings would produce different quantum effects, and the potentials (\ref{corda1}) and (\ref{corda2}) could not describe the same physical system, the magnetic monopole.

	After Dirac's seminal paper on magnetic monopoles, there appeared several others in the literature \cite{GoddardOliverRPP78,MignacoBJP01}, all of them presenting quantization conditions for the electric charge, but involving magnetic monopoles as well.
	
	We aim here at reassessing charge quantization in the absence of monopoles. To do that, inspired by the Dirac's gauge transformation (\ref{calibreDirac}), let us consider the peculiar gauge transformation
\begin{equation}
\label{calibrenosso}
{\bf A}\rightarrow{\bf A}'={\bf A}+{\bf A}_{\kappa}\ ,\ {\bf A}_{\kappa}={\bf\nabla}(\kappa\phi)=\frac{\kappa}{\rho}{\hat\phi}\ ,
\end{equation}
where we have the cilindrical coordinate $\rho=r\sin(\theta)$ and $\kappa$ is a constant. Note that, as (\ref{calibreDirac}), the transformation (\ref{calibrenosso}) must be taken only in a patch which excludes the $z$-axis.

	In modern physics gauge transformations are strong statements, therefore, it is consistent to consider that a given gauge transformation can always be performed, unless there is a reason for not doing it. Now, what we have to do is to investigate if the gauge transformation (\ref{calibrenosso}) can be performed, but before this, let us make some remarks on the potential ${\bf A}_{\kappa}$ of (\ref{calibrenosso}).
	
	We would like to stress on the fact that the potential (\ref{calibrenosso}) is not in desagreement with the Maxwell Electrodynamics, once its rotational is equal to zero in the domain where it is defined\cite{footnote}. Consequently, the flux of its rotational through any surface setted on its domain patch is also zero. As stated in reference\cite{WC2004}, the transformation (\ref{calibrenosso}) is not any longer in desagreement with Stokes' Theorem and with the Aharonov-Bohm poblem, as will be discussed in this paper.
	
	In what concerns the magnetic fiel flux and the Stoke's Theorem, we have to consider two cases separately. In the first case, we have the contribution of ${\bf A}_{\kappa}$ in (\ref{calibrenosso}) to the magnetic flux through any surface which does not intercepts the $z$-axis. A surface of this kind is bounded by a loop which does not encircles the $z$-axis. Noting that the integral of ${\bf A}_{\kappa}$ along any loop of this kind is always zero, it is easy apply the Stoke's Theorem to (\ref{calibrenosso}) and conclude that it is zero.
	
	The second case is where we have a surface which intecepts the $z$-axis. To simplify the problem, we shall consider the situations where this interception happens only once. In this case the boundary of the surface encircles the $z$-axis and the integral of ${\bf A}_{\kappa}$ along any loop of this kind is equal to $2\pi\kappa$. To apply Stokes' Theorem to this case it is important to note that the contribution of ${\bf A}_{\kappa}$ to the magnetic flux through the considered region comes from the overlap between this region and the patch where the potential ${\bf A}_{\kappa}$ is defined, which excludes the $z$-axis, it is done as follows
\begin{eqnarray}
\int\int d{\bf S}\cdot{\bf\nabla}_{\times}{\bf A}_{\kappa}&=&\oint_{P}d_{\bf\ell}\cdot{\bf A}_{\kappa}-\oint_{p}d_{\bf\ell}\cdot{\bf A}_{\kappa}\cr\cr
&=&2\pi\kappa-2\pi\kappa=0
\end{eqnarray}
where $\bf S$ is the normal to the surface, $P$ designates the boundary of the surface and $p$ is an infinitesimal loop on the surface, which encircles the $z$-axis.

	Therefore, for any situation, the potential ${\bf A}_{\kappa}$ gives no contribution to the flux of magnetic field.
	
	Now, let us analyze if we can perform the transformation (\ref{calibrenosso}).
	
	As already pointed out, the transformation (\ref{calibrenosso}) has the same structure of (\ref{calibreDirac}), and by the same reasons which lead to equation (\ref{quantizacaoDirac}), in order to have the potential ${\bf A}_{\gamma}$ as a pure gauge one, we must have the condition $q\kappa=n_{q,\kappa}$ where $q$ is the value of electric charge that a wave function can have and $n_{q,\kappa}$ is an integer which depends on the parameter $\kappa$ and the charge $q$. With this condition it is stated that the potential ${\bf A}_{\gamma}$ does not produce any quantum physical effect.
	
		In order to ensure the independence between the charge $q$ and $\kappa$, the latter should be considered merely as a gauge parameter. We can write, without loss of generality,
\begin{equation}
\label{condicaokappa2}
q\kappa=n_{q}n_{\kappa}\ ,
\end{equation}
where $n_{q}$ and $n_{\kappa}$ are integers that depend respectively on the charge $q$ and the parameter $\kappa$. Using in equation (\ref{condicaokappa2}) the charge of the electron $e$, for example, we are taken to
\begin{equation}
\label{condicaokappa3}
\kappa=\frac{n_{e}n_{\kappa}}{e}=\frac{Nn_{\kappa}}{e}\ ,
\end{equation}
where we defined $N=n_{e}$ which is a non-zero integer constant. Replacing equation (\ref{condicaokappa3}) into (\ref{condicaokappa2}), we have
\begin{equation}
\label{resultado}
q={n_{q}\over N}e\ ,
\end{equation}
i.e., any value of the electric charge $q$ is an integer multiple of a given fraction of the electron charge $e$.

	It is interesting to notice that the result (\ref{resultado}) is consistent with the fractional values of the quark charges and with the existence of anti-particles, since, for each value of $q$, this relation predicts also the value $-q$ as another possibility.

	From equation (\ref{condicaokappa3}), we can see that the possible values of the $\kappa$ parameter, that, we stress, was completely free in the classical level, becomes quantized in the quantum context.

	Now, let us investigate the implications of the gauge transformation (\ref{calibrenosso}) to the Aharonov-Bohm effect.

	In their well-known papers of 1959 and 1961, Aharonov and Bohm \cite{AB,revAB} considered an infinitely long solenoid inside which we could have a uniform and constant magnetic field; outside no magnetic field is present.

	The magnetic field generated by an infinitely long solenoid of radius $R$, with its axis lying on the $z$ axis, is given, in cylindrical coordinates, by
\begin{eqnarray}
\label{defB}
{\bf B}_{sol}(\rho,\phi,z)&=&\cases{{\bf B}_{I}=B{\hat z},&$\rho<R$\cr {\bf B}_{E}={\bf 0},&$\rho>R$}\cr\cr
&=&B\Theta(R-\rho){\hat z}\ .
\end{eqnarray}	
On the solenoid, $\rho=R$, the magnetic field is not defined; it is a singularity region.

	In order to describe the field (\ref{defB}) Aharonov and Bohm\cite{AB,revAB} used the vector potential
\begin{eqnarray}
\label{defAAB}
{\bf A}(\rho,z,\phi)&=&\cases{\bigl(B\rho/2\bigr){\hat\phi}, \ \rho<R\cr\cr \bigl(BR^{2}/2\rho\bigr){\hat\phi}, \ \rho>R}\cr\cr\cr
&=&\frac{B}{2}\Biggl[\rho\Theta(R-\rho)+\frac{R^{2}}{\rho}\Theta(\rho-R)\Biggr]{\bf\hat\phi}\ .
\end{eqnarray}
Notice that, outside the solenoid, $\rho>R$, the potential (\ref{defAAB}) has the same structure as the Dirac's gauge field (\ref{calibreDirac}) (That is why we called the field ${\bf A}_{\kappa}$ in (\ref{calibrenosso}) as Dirac-Aharonov-Bohm potential).	Using simple results on distributions, the vector potential (\ref{defAAB}) gives the magnetic field (\ref{defB}).
	
	As stated by Aharonov and Bohm\cite{AB,revAB}, in spite of the vector potential (\ref{defAAB}) have rotational equal to zero in the exterior region of the solenoid, it can produce, in the context of Quantum Mechanics, physical phenomena in this region, called usually as Aharonov-Bohm effects. 

	In the Aharonov-Bohm problem, once the domain of the wave function excludes the $z$-axis, we have a more confortable situation to perform the gauge transformation (\ref{calibrenosso}). Besides, it is more natural to do this transformation only on the exterior region of the solenoid ($\rho>R$).
	
	In addition to this approach for modifying the Aharonov-Bohm potential (\ref{defAAB}), we could add to (\ref{defAAB}) the potential 
\begin{equation}
\label{potencial2}
{\bf\cal A}=\frac{\kappa}{\rho}\Theta(\rho-R){\hat\phi}\ ,
\end{equation}
which is defined in the whole space except on the solenoid $\rho=R$.

	The potencial (\ref{potencial2}) produces the contribution
\begin{equation}
{\bf\cal B}={\nabla}_{\times}{\bf\cal A}=\frac{\kappa}{\rho}\delta(\rho-R){\hat z}
\end{equation}
to the magnetic field which exists only on the solenoid. Once the solenoid is not a physical region, we can not state what is the magnetic field on it, even more, it is a surface current density, and obtain a divergent magnetic field on it is not properly a surprise.
All we can state about the field strength are their properties on the physical regions, and these ones are inalterated by the addition of (\ref{potencial2}). Besides, the addition of (\ref{potencial2}) does not produce any physical effect once $\kappa$ is restricte by (\ref{condicaokappa2}).

	Some years ago, the gauge transformation (\ref{calibrenosso}) was, erroneously, used to argue on the non-existence of the Aharonov-Bohm effect \cite{revAB,1}, and some times, these arguments were refused with remarks based on the Stokes' Theorem \cite{revAB,2}. As already stressed in the work of reference \cite{WC2004}, we would like to say that a transformation of the form (\ref{calibrenosso}), or the addition of (\ref{potencial2}) in the case of solenoid, are not in disagreement with Stokes' Theorem. So, the final outcome of our discussion is that the reassessment of the problem of a magnetig Dirac-Aharonov-Bohm-like vector potential taken as a gauge field, leads us to the interesting possibility of getting charge quantization, without the usual argument based on the existence of a magnetic monopole.

	These results could be extended to the case of non-Abelian fields, where a quantization condition for the color charges must be obtained too \cite{naoabeliano}.

The authors would like to thank M. V. Cougo-Pinto, C. Farina, F.C. Santos, H. Boschi-Filho and F.E. Barone for useful discussions. Professor R. Jackiw is kindly acknowledged for relevant comments, and Professor Nathan Lepora for the help with relevant references. F.A.B. thanks FAPERJ for the invaluable financial help.




\begin{thebibliography}{*}
\bibitem{footnote} Although the potential (\ref{calibrenosso}) can produce a divergent rotational on the $z$-axis, it is excluded from the patch where the potential is defined.
\end{thebibliography}
\end{document}